# Emerging Results on Automated Support for Searching and Selecting Evidence for Systematic Literature Review Updates

Bianca Minetto Napoleão
Université du Québec à Chicoutimi
Chicoutimi, Canada
bianca.minetto-napoleao1@uqac.ca

Ritika Sarkar
Université du Québec à Chicoutimi
Chicoutimi, Canada
rsarkar@etu.uqac.ca

Sylvain Hallé
Université du Québec à Chicoutimi
Chicoutimi, Canada
shalle@acm.org

Fabio Petrillo
École de Technologie Supérieure
Montreal, Canada
fabio.petrillo@etsmtl.ca

Marcos Kalinowski
Pontifical Catholic University of Rio de Janeiro
Rio de Janeiro, Brazil
kalinowski@inf.puc-rio.br

## ABSTRACT

**Context:** The constant growth of primary evidence and Systematic Literature Reviews (SLRs) publications in the Software Engineering (SE) field leads to the need for SLR Updates. However, searching and selecting evidence for SLR updates demands significant effort from SE researchers. **Objective:** We present emerging results on an automated approach to support searching and selecting studies for SLR updates in SE. **Method:** We developed an automated tool prototype to perform the snowballing search technique and to support the selection of relevant studies for SLR updates using Machine Learning (ML) algorithms. We evaluated our automation proposition through a small-scale evaluation with a reliable dataset from an SLR replication and its update. **Results:** Effectively automating snowballing-based search strategies showed feasibility with minor losses, specifically related to papers without Digital Object Identifier (DOI). The ML algorithm giving the highest performance to select studies for SLR updates was Linear Support Vector Machine with approximately 74% recall and 15% precision. The use of such algorithms with conservative thresholds to minimize the risk of missing papers can already significantly reduce evidence selection efforts. **Conclusion:** The preliminary results of our evaluation point in promising directions, indicating the potential of automating snowballing search efforts and of reducing the number of papers to be manually analyzed by about 2.5 times when selecting evidence for updating SLRs in SE.

## 1 INTRODUCTION

Evidence-Based Software Engineering (EBSE) has as the main instrument Systematic Literature Reviews (SLRs) to summarize evidence from primary studies (e.g., case studies, surveys, controlled experiments), providing recommendations for practitioners and supporting the decision-making process in Software Engineering (SE) [15].

Over the almost 20 years of SLRs in SE, the number of SLRs has increased substantially [17, 19], leading to the need to update SE SLRs [17]. A not-maintained SLR could lead researchers to obsolete conclusions or decisions about a research topic [33] (e.g., studies that employed a particular technology that is now outdated) [36].

Performing SLR updates demands significant effort especially because of the rapid increase of available evidence [31, 38]. Identifying studies for SLR consists of two main activities: (i) searching for potentially relevant studies and (ii) selecting the real relevant ones.

Aiming to facilitate the search activity, Felizardo *et al.* [8] introduced the use of forward snowballing (i.e., citations analysis from the included studies in the original SLR a.k.a. seed set [34]) to support searching for evidence in SLR updates. A few years later, Wohlin *et al.* [36] further evaluated search strategies for SLR updates and found that the use of a single iteration forward snowballing employing as a seed set the original SLR and its primary studies tends to be the most cost-effective way to search for new evidence when updating SLRs.

With respect to reducing the effort of study selection considering specifically the SLR update scenario, only two attempts have been investigated by researchers [11, 33]. Both studies present approaches addressing automation support: Felizardo *et al.* [11] based their approach on Visual Text Mining (VTM) techniques, while Watanabe *et al.* [33] demonstrated the potential of supervised Machine Learning (ML) algorithms to support the selection activity.

The goal of this study is to propose and evaluate an automated an approach to support both the searching and selecting of studies for SLR updates in SE. To achieve this goal, we developed an automated tool prototype to perform the snowballing search technique [34] and also to support the selection of relevant studies for SLR updates by investigating the practical use of ML algorithms. We evaluated our automation proposition by performing small-scale evaluation [37] with a reliable dataset from an SLR replication [35] and an ongoing update. Our preliminary results indicate that snowballing-based search strategies can be fully automated with minor losses and that significant effort can be saved in SLR update study selection using ML, allowing a conservatively reduced number of papers to be manually analyzed by about 2.5 times.

The remainder of this study is organized as follows. Section 2 presents a brief background and related work. In Section 3 we describe the prototype tool. The small-scale evaluation is reported in Section 4. Section 5 discusses our results. Section 6 relates threats to validity. Finally, Section 7 concludes our work.

## 2 BACKGROUND AND RELATED WORK

According to Mendes *et al.* [17] an SLR update is a new version of a published SLR that includes new primary studies that can come from a different search strategy than the original SLR (e.g., snowballing, manual or database search). Identifying new relevant evidence for updating an SLR is one of the initial steps in identifying the need and possibility of updating an SLR.

In SE, there are studies that investigated search strategies dedicated to searching for evidence for updating SLRs. Felizardo *et al.* [8] proposed the use of forward snowballing as a search strategy to update SE SLRs. Their results showed a reduction of more than five times the quantity of primary studies to be analyzed during an SLR update. Later, in 2020, Wohlin *et al.* [36] further investigated the use of forward snowballing to update the SE SLRs. Their study proposed and evaluated guidelines for the search strategy to update SLRs in SE. They found that using a single iteration of forward snowballing, with Google Scholar as the search engine, and employing the original SLR and its primary studies as a seed set tends to be the most cost-effective way to search for new evidence for updating an SLR. One of the goals of our study is to automate forward snowballing as proposed by [8, 36] to facilitate the search activity.

Given that the selection of evidence resulting from the execution of the search strategy is a laborious activity [7, 10], there are two related works that address alternatives to remedy this issue in SLR updates. The first one by Felizardo *et al.* [11] explored visual text mining to support selecting new evidence (primary studies) for SLR updates. The tool presented, called Revis, connects the new evidence with the evidence of the original SLR applying the KNN (K-Nearest Neighbor) Edges Connection technique presenting the results in two different visualizations: content-map and Edge Bundles diagram. The results showed an increase in the number of studies correctly included compared to the traditional manual approach.

The second one by Watanabe *et al.* [33] also takes advantage of the fact that a published SLR that needs updating already has a list of included studies. They investigated using text classification techniques, including supervised ML algorithms (Decision Tree and Support Vector Machines), to make the initial selection of primary evidence (based only on the title and abstract of the studies) to update SLRs. The study indicated potential of using automated techniques to reduce the effort required to select studies for SLR updates.

It is worth mentioning that automation of the search and selection of studies has been also investigated in the context of SLR conduction, for example: (i) Carver & Felizardo [9] present an overview of existing automation alternatives for all the activities of the SLR process; and (ii) Napoleão *et al.* [20] focus on the search and selection activities presenting a systematic mapping on existing automation support to search and selection of studies addressing both the SE and medicine domains.

Our study builds on knowledge gathered in the related work and proposes and investigates a search and selection approach with main focus on the SLR update scenario.

## 3 TOOL DEVELOPMENT

In this Section, we present details about our prototype tool developed to support the snowballing search activity and the selection of studies activity for SLR Updates.

We started with the development of a prototype tool to perform both snowballing techniques, forward (citation analysis) and backward (references analysis) [34]. Even though the main objective of this study is to investigate automation support for searching and selecting studies for the SLR updates, which do not require backward snowballing, during development we noticed that the design of the forward solution was easily replicated for the backward solution. Therefore, we opted for the development of a snowballing tool addressing both types.

The execution flow of our proposed algorithm for developing the snowballing prototype tool is shown in Figure 1. The snowballing automation is preceded by inputs from the user in the form of Digital Object Identifier (DOI) URLs of papers in the seed set. Following this, the implementation code is run and the user is asked to inform the number of snowballing iterations he/she wants to run and whether they wish to proceed with either backward or forward snowballing, or both. We chose to add these two inputs because, for the context of SLR updates, a single iteration of forward snowballing is enough to return the relevant studies [8, 36].

The snowballing solution (backward and forward) is implemented by querying the Semantic Scholar API [30] based on the DOI of studies. The metadata returned as the query results are employed to extract the DOIs of citations and references cited in the queried study. In case of studies without DOIs, CrossRef API [5] is queried for DOIs by providing the input as a reference string generated using the keys 'authors', 'title', 'venue', and 'year' returned in the metadata by Semantic Scholar [30].

Next, the acquired DOIs are passed through a redundancy check (for subsequent iterations) to ensure that the extraction has not been done for them in previous iterations. The main part of the data acquisition starts with getting the bibliographical metadata of the references and citations in the BibTeX format, by making a request to the DOI using the Urllib [24] library.

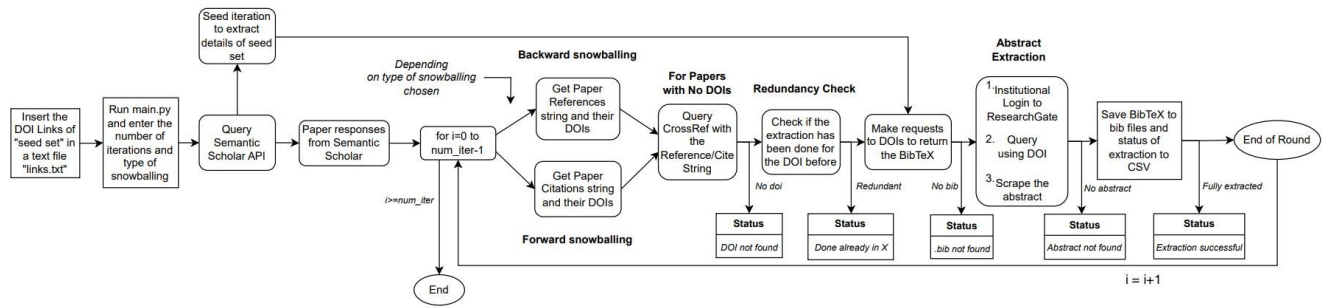

Figure 1: Execution flow of the snowballing tool

The abstract is usually missing in the response and hence we need to apply other methods to get the abstract. For the abstract extraction, we make use of ResearchGate [27], which allows researchers and students free access to abstracts of scientific publications and preprints of research work. We employed web scraping to log in to ResearchGate with institutional credentials, and query for the studies using the DOIs. As the page is rendered by the browser, the abstract is extracted and saved to the BibTeX of the corresponding study.

The results of the multiple iterations of the snowballing process are stored in one common CSV file and one common BibTeX file at the end of the runs. This applies to each type of snowballing. Therefore, if the user wishes to perform both backward and forward snowballing together, there will be 2 separate files (CSV and .bib) for each type of snowballing. The CSV file stores the reference strings, the corresponding DOIs, the status of the extraction, and the iteration number. The reference string is generated using the keys 'authors', 'title', 'venue', and 'year' returned in the metadata by Semantic Scholar [30], which are joined to produce a string similar to the Chicago format of referencing. This is done because the full reference of a study is not returned by the Semantic Scholar. The status of the extraction for a particular study uses the following implicit phrases: "Extraction successful", "DOI not found", ".bib file not found", "Abstract not found", and "Done already in X" where X is the iteration number.

Since we store the results of all the iterations in a common CSV file, the column "iteration number" tells us in which iteration of snowballing a particular study was discovered. The BibTeXs of the studies are stored in a common .bib file. Each BibTeX is appended to the common BibTeX file after each extraction phase. Another feature of the tool helps us to obtain the BibTeX file for the seed set as well.

Regarding the selection of studies portion of the tool, we opted to evaluate a range of ML algorithms known to perform well for text classification [1, 22]: XGBoost (XGB) [4], Linear Support Vector Machines (LSVM) [16], Logistic Regression (LogReg) [12], and Multinomial Naïve Bayes (MNB) [14]. We chose them because all four algorithms have a regularization term, which plays an important role in combating overfitting and underfitting in unbalanced datasets by adding penalties to the loss function. The second-most important term is "class weight" which prevents the prejudice of the model towards the minority class, by assigning a higher weight to it. They are reciprocal of the class frequencies. We detail the tool's selection process and parametrization in Section 4.2 by providing a practical evaluation example.

## 4 SMALL-SCALE EVALUATION

To evaluate our prototype tool, we performed a small-scale evaluation [37]. According to the smell indicator proposed by Wohlin & Rainer [37], the correct label for our evaluation is small-scale evaluation instead of a case study. However, to guide and report our evaluation process we followed the five main steps for case studies proposed by Runeson *et al.* [28]: Design, preparation, collecting data, analysis, and reporting.

### 4.1 Design

Our design consists in selecting a published SLR replication [35] and its ongoing SLR update evaluation instrument to search for studies through snowballing iterations and perform an initial selection of potentially relevant studies to be included in an SLR update.

To evaluate the prototype tool's capability of performing backward and forward snowballing, we opted to use the SLR replication study [35] since it documents in the supplementary material[1] the results of each snowballing iteration. In summary, our goal with this is to illustrate the tool's potential to be used to perform both snowballing search types in an SLR conduction process when a seed set of studies is known by the authors (e.g. selected from database search [35]).

Next, to evaluate the tool's capability of being employed in the SLR update context, the main goal of our study, we used the 45 selected studies by the SLR replication [35] as a seed set to perform an iteration of forward snowballing and then apply the ML algorithms on a reliable and complete dataset from the ongoing SLR update of [35]. In this replication and the ongoing update, the inclusion and exclusion of new studies were conducted based on individual assessments and the consensus of three experienced SLR researchers, allowing us to have confidence in this data for building reliable training and testing sets.

### 4.2 Preparation and Collecting data

To evaluate the tool's capabilities to perform backward and forward snowballing we first prepared our seed set by obtaining the DOI of the studies mentioned as seed set (9 studies) in the supplementary data of the SLR replication [35]. Next, we replicated with our tool the same 5 iterations of backward and forward snowballing performed manually by the authors. Finally, we compared our tool's results with the manual execution.

---

[1] https://ars.els-cdn.com/content/image/1-s2.0-S0950584922000659-mmc2.pdf

Regarding the analysis of the search and selection for SLR updates, we conducted three steps. First, we search performing a single forward snowballing iteration using the 45 included studies by the SLR replication [35] as seed set. Second, we train our ML algorithms with the "training set" containing both included and excluded studies of the SLR replication [35]. Finally, we perform the selection of studies using the trained algorithms on the results of the forward snowballing in the first step ("testing set").

The distribution of included and excluded studies in the training and the testing sets are shown in Figure 2. It is highly imbalanced with a minority of included studies. The imbalance represents the real-world scenario where out of a large number of studies only a few are typically relevant to the focus of a particular research topic.

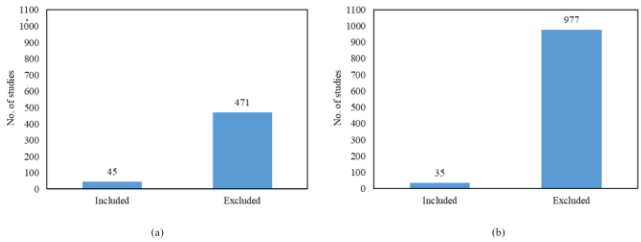

**Figure 2: The data distribution of (a) Training set and (b) Testing set**

The results obtained using our algorithms are compared with the included articles selected manually for this SLR update (the "testing set") to evaluate the recall, precision, and F-measure of the selected algorithms and the automated snowballing tool. Figure 3 summarizes the evaluation of the tool's selection process.

Regarding the training process, for *Training Data Collection*, as shown in Figure 3, the input consists of the included and excluded studies in BibTeX format (.bib) of the SLR replication [35]. We converted the .bib file into CSV format by taking the 'Title' and 'Abstract' fields of the studies. We also label them with relevance 1 for included studies in the SLR replication [35], and relevance 0 for excluded studies from the SLR replication [35].

Thus the labeled Training set CSV is ready to be passed to the *Preprocessing* phase. The block *Training* shows the training process after the training set is fed to the Binary Classifiers LSVM, XGBoost, LogReg and MNB, for training. After the initial training, hyperparameter tuning is done using the different parameters of the models to improve their performance on the minority class (here, 1). Hyperparameter tuning was done by rerunning the algorithms several times with different values to find the model parameters most suited to our goal: maximizing recall (finding most of the relevant studies) [15] and precision (reducing the load on reviewers to check irrelevant studies). Our goal is to get at least an acceptable trade-off between recall and precision according to the classification presented in [6]. The best hyperparameter configurations obtained during the training phase using Sklearn toolkit [29] were as follows: the hyperparameter term "alpha" is set as 2 to perform strong regularization on the LSVM model, and both classes are given due importance by setting the "class_weight" term as 'balanced' which follows a weighted loss function. This linear model is then trained using Stochastic Gradient Descent (SGD) [3] to optimize the loss function with a decreasing learning rate. A similar approach was followed for XGBoost and Logistic Regression. In XGBoost we set "gamma" (regularization term) as 20, the "scale_pos_weight" term as (number of articles in class 0)/(number of articles in class 1), and "sub_sampling" ratio term to 0.2 to prevent overfitting. In Logistic Regression, "C" (the regularization term) is set to 0.01 and the "class_weight" term is set to 'balanced'. For Multinomial Naïve Bayes the default parameters of Sklearn are used. In the first three models, strong regularization was done to maximize the recall and precision of the minority class by making the models more conservative and generalize better on testing data.

Concerning the testing process, *Testing Data Collection* is divided into two parts. First, *Testing Data Collection Part 1* implements one round of forward snowballing using the snowballing tool on the update seed set (45 selected studies from the SLR replication). The output of the forward snowballing process is a CSV file keeping track of the study extraction and a BibTeX file which holds the bibliographical references of all the studies including their abstracts, in order to aid the authors in the selection of relevant studies. This task which is usually done manually is completely automated. Similarly as done for the training, the BibTeX file from the output of forward snowballing is converted to a CSV file taking the 'Title' and 'Abstract' fields of the studies. Second, in *Testing Data Collection Part 2* the BibTeX file of included papers is also converted to CSV and a comparison is done between the two CSVs to generate a unique labeled CSV file for testing, labeling the relevance of included studies as 1 and excluded ones as 0.

Thereafter, the labeled testing set CSV is passed through the *Preprocessing* phase. Finally, during *Testing* the trained model is used to predict inclusion or exclusion for the preprocessed testing data.

*Preprocessing* is done similarly for the training and test CSV files. First, the 'Title' and 'Abstract' columns are merged to form a single string for each study under a column named 'Merged', which now serves as the text for text classification. The preprocessing treats this text by removing stop-words using the Nltk [21] library in Python [18], tokenizing the text, removing punctuation and URLs, performing lemmatization [23], and finally vectorizing using the Bag of Words [32] count vectorizer. The count vectorizer gives poor results when the n-gram range is increased as it predicts with a greater precision only for the majority class. In this sense, unigrams are used. The preprocessed text is then passed as input to the ML algorithm models for training.

Finally, for *Validation*, we record the number of included studies identified during the forward snowballing round. The labels of the test set allowed us to compute the performance measures Precision, Recall, and F-measure based on the predictions obtained during the *Testing* phase. They are defined as follows [15]:

$$recall = \frac{number\ of\ relevant\ studies\ retrieved\ as\ relevant}{total\ number\ of\ relevant\ studies}$$

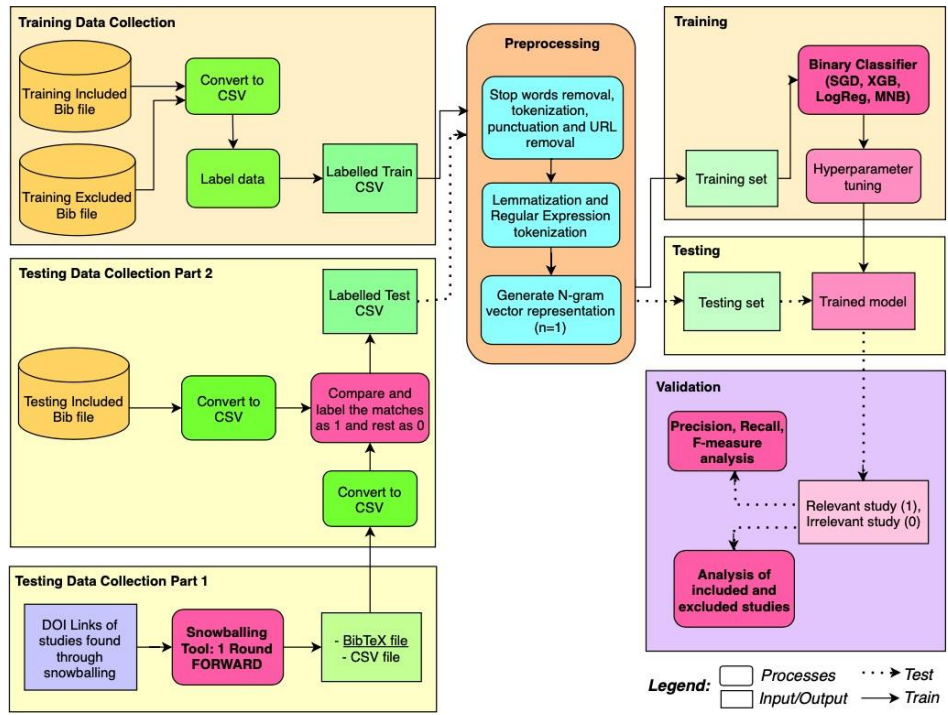

**Figure 3: Tool process to select studies for SLR Updates**

$$precision = \frac{number\ of\ relevant\ studies\ retrieved\ as\ relevant}{total\ number\ of\ studies\ retrieved\ as\ relevant}$$

*F-measure* = *harmonic mean of the precision and recall*

The results of our evaluation are presented in Section 4.3.

## 4.3 Results - Analysis of our evaluation

The evaluation of our tool prototype for the reproduction of backward and forward snowballing for the SLR replication is presented in Table 1. In total, our tool was able to automatically identify 41 of the 43 (95.3%) studies manually identified by the authors when performing the same snowballing iterations (see Table 1 sum of column "Studies detected" + nine studies from the initial seed set). However, during iteration 2, a study identified manually during forward snowballing was identified by our tool during the backward snowballing execution (see lines highlighted with (*) in Table 1). We opted to conserve this result since it does not interfere with the final result of the snowballing tool in this automated execution scenario.

**Table 1: Results of the snowballing search replicated by our tool**

| Snowballing type | Seed set of the iteration | Studies detected (%) |
|---|---|---|
| **Iteration 1** | | |
| Backward | 9 | 12/12 (100%) |
| Forward | 9 | 1/1 (100%) |
| **Iteration 2** | | |
| Backward | 13 (1+12) | 1+1* = 2/1 (100%) |
| Forward | 13 (1+12) | 12/14 (85.7%) |
| **Iteration 3** | | |
| Backward | 14 (2*+12) | 3/4 (75%) |
| Forward | 14 (2*+12) | 1/1 (100%) |
| **Iteration 4** | | |
| Backward | 4 (3+1) | 1/1 (100%) |
| Forward | 4 (3+1) | 0/0 |

The two missing studies could not be identified because they do not have a DOI and the tool was not able to locate them through the implemented reference building (Chicago format). It is worth mentioning that the set of included studies of the SLR replication is composed of 45 studies in total, but two of them are not considered in this analysis because they were not retrieved by snowballing. In addition, our tool was able to execute automatically four iterations instead of five (manual) because the single study resulting from iteration 4 did not have a DOI available which made the snowballing process stop. However, in this case, the final result is still the same (manual versus automated) since the manual process also stopped for not retrieving any other relevant study.

Regarding the evaluation of our tool applied in an SLR update scenario, the snowballing tool was able to retrieve 1012 unique studies in a single round of forward snowballing based on a seed set of 41 out of 45 studies included in the SLR replication. The 4 remaining studies not identified did not have DOIs, consequently, we missed out on 28 citations (data from Google Scholar in Feb. 2023). This leads to our final seed set being formed by the 41 studies and having 1012 unique citations to be analyzed in the selection phase by the ML algorithms.

Out of 35 studies contained in the "testing set - included", that are to be included in the SLR update, our search and selection tool identified 33 studies (94.3%) through the forward snowballing iteration, even without being able to include the 4 no-DOI studies in the seed set.

The precision, recall, and F-measure scores on the testing for the class of interest (included papers) are illustrated in Figure 4. It can be seen that the performing model in terms of recall is LSVM (74.3%), followed by XGBoost (63.6%), Logistic Regression (45.4%), and Multinomial Naïve Bayes (42.4%). In terms of precision, the algorithms had almost similar results (~15%) with the exception of XGBoost which had a lower precision (11.6%). Considering the fact that the F-measure combines the effect of both metrics, a high recall will give a low F-measure if the precision is low. The F-measure value was observed with the LSVM model (24.6%). It is able to predict the highest number of studies belonging to the positive class 1 (included), correctly. The precision value can be explained by the high false positive value which is due to the strict regularization performed in LSVM, hence, a trade-off between precision and recall is remarked. In fact, Dieste *et al.* [6] highlight that there will always be a trade-off between recall and precision because irrelevant studies are more liked to be returned by a search execution the higher the recall is.

Out of 33 studies, 26 studies were identified by the LSVM model, 21 by XGB, 15 by LogReg, and 14 by MNB. The studies to be excluded correctly identified were 820 by LSVM and XGB, 890 by LogReg, and 900 by MNB. LogReg and MNB give biased results for the majority class hence the true negative values are much higher than LSVM, but they have lower true positive values which is the number of included studies predicted correctly by the model.

### 4.4 Reporting - Observations from the evaluation

According to the analysis of the performance measures, the LSVM model showed the best result among the evaluated models. Moreover, following the search strategies scale proposed by Dieste *et al.* [6], the recall and precision range resulted from the LSVM model showed to be acceptable (recall 72-80% and precision 15-25%) meaning a "good enough strategy".

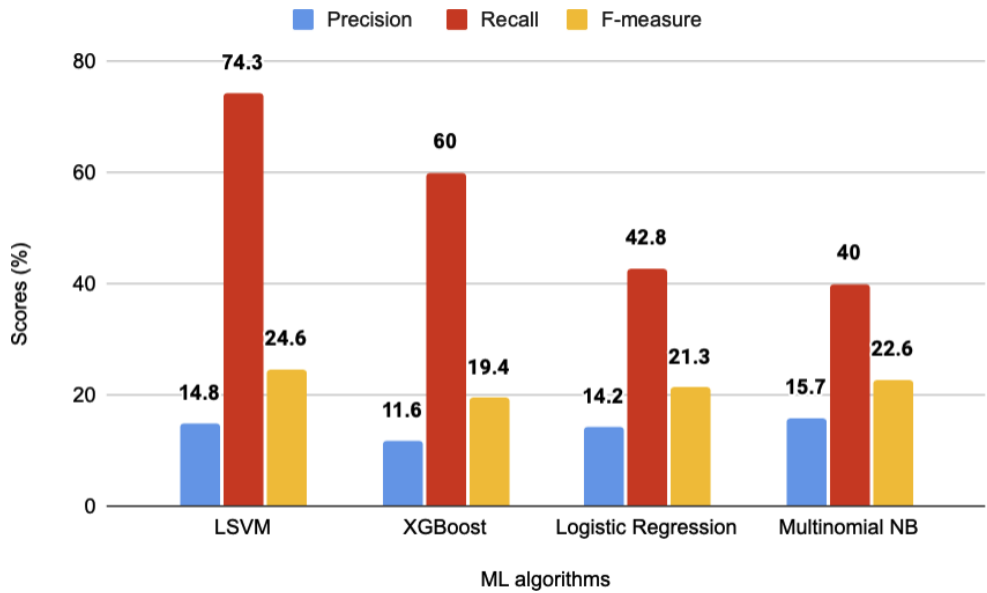

**Figure 4: The performance report of the ML models**

## 5 DISCUSSION

The motivation of this study was to investigate automated alternatives to support both activities of searching and selecting studies during SLR updates. Due to the effort required for these activities, authors often end up not giving enough attention to keeping SLRs up to date. The impact of not updating SLRs on the evolution of scientific knowledge can be severe, as outdated SLRs could lead researchers to obsolete conclusions. While we present emerging results, they already shed light on some meaningful automation limitations and possibilities.

With respect to searching for new studies, we described solution options used when building our snowballing tool to face automation challenges (e.g., querying CrossRef to find DOIs, and complementing BibTex data with web scrapping). Besides providing an example of such solution options, our results indicate that snowballing-based search strategies can be fully automated with minor losses, at least for classical white literature (the only identified limitation was related to papers without DOI), reducing the effort of laborious manual snowballing iterations (*cf.* Table 1). In order to mitigate minor losses and the potential inclusion of non-classical white literature, manual searches [15] can be performed to complement the results of the snowballing search.

Regarding the selection of studies in SLR updates using ML algorithms trained based on papers included and excluded in the original SLR, the recall and precision obtained by these algorithms still represent only a "good enough strategy", according to the scale proposed by Dieste *et al.*. Still, they would allow to conservatively save some manual selection effort. In fact, if we take our proposed model (LSVM), even if we use a threshold that results in a max recall scenario of 97% for the classifier (one false negative) and only 8% of precision, the SLR update author would have to manually analyze only 396 papers instead of 1012. *I.e.*, for our investigation, in a scenario of conservatively minimizing the risk of missing papers to be included, it would still be possible to reduce the number of papers to be manually analyzed by more than 2.5 times.

Overall, we believe that our results provide preliminary indications that strengthen the belief that automated approaches could significantly help to reduce the SLR update efforts.

Large Language Models (LLMs) gained attention over the last year. The main difference between LLMs and traditional ML algorithms combined with Natural Language Processing techniques is that ML algorithms often rely on labelled or annotated datasets specific to the SLR task. These datasets are typically created manually by domain experts. On the other hand, LLMs are pre-trained on large general-domain unannotated datasets, and their language understanding capabilities can be fine-tuned with smaller domain-specific datasets (e.g. specific domain requirements of an SLR) [25, 39]. One may question the possible application of LLMs in the context of selection of studies for SLRs or SLRs updates. To the best of our knowledge, there is an initiative [2] that explored a deep-learning-based contextualized embedding clustering technique employing two transformer-based language models BERT [13] and S-BERT [26] to perform the initial selection of studies for SLRs. To evaluate their proposal, the authors compared the generated models' resulting clusters with the results of two SLRs from medicine manually conducted. As a result, due to the small size of the training set, the models ran out of data to train and suffered from overfitting. The authors affirm that an extension with a larger dataset is needed to underline a conclusion [2]. In summary, LLM is still an emerging topic and requires further investigation and validation before being applied on the context of selection of studies for SLRs and SLRs updates.

## 6 THREATS TO VALIDITY

In the following, we enumerate the main threats to the validity of our study.

**Construct Validity.** With respect to searching for studies using snowballing, the adoption of Semantic Scholar has not been formally evaluated by researchers in the context of SLR updates [36], as Google Scholar. However, we noticed that, in our case, the results showed to be relevant for the study and comparable, with less noise. Also, Google Scholar does not allow the use of an API to perform searches. For study selection, our evaluation results might have been affected by the choice of ML algorithms. Other algorithms could have been explored in our study and can be considered as part of future work.

**External Validity.** The dataset used in our analysis might not represent the diversity of SLR Updates in SE. Similar analyses could have been conducted based on other SLRs to improve the generalizability of our results. However, replicating our emerging results on other SLRs to strengthen external validity would require significant effort. Furthermore, it is challenging to acquire a reliable and detailed SLR dataset (e.g. containing the list of included studies in each iteration of the snowballing search) for SLRs that could potentially need an update and be considered in our analysis.

**Reliability.** One limitation of our study is associated with the dataset used in our experiment and the possibility of sample bias. For the snowballing analysis, we used a dataset of an SLR replication that involved experienced SLR researchers following strict guidelines for searching and selecting evidence [35]. The data used for the SLR Update analysis was acquired from the same authors who performed the SLR replication, also through a rigorous analysis process. Also, to improve the reliability of our results, the tool prototype and the small-scale evaluation datasets are openly available.

## 7 CONCLUSION

In this paper, we presented and investigated an automation solution proposal to support searching for new evidence and selecting evidence for SLR updates. We built a tool prototype and described it in detail. Based on a small-scale evaluation, we discuss automation limitations and perspectives for the SLR update context.

Concerning searching for evidence, preliminary results of our investigation indicate that, while there are challenges faced when automating snowballing-based search strategies (e.g., to automatically gather DOIs for papers, to complement BibTeX information of identified papers automatically), these strategies can be fully automated with minor losses. This can be particularly helpful for updating SLRs, given that forward snowballing has been recommended for this context [8, 36]. Furthermore, applying automated snowballing iterations could also be employed to reduce the effort of applying SLR search strategies in general [35].

We also investigated the selection of studies in SLR updates using ML algorithms trained based on papers included and excluded in the original SLR. While improvements can surely be obtained, emerging results obtained by our prototype tool are promising and already considered acceptable by literature [6]. In our small-scale evaluation, it was also possible to observe that using our proposed investigation using ML model (Linear SVM optimized using the SGD algorithm) conservatively minimizing the risk of missing papers during the SLR update, it would still be possible to reduce the number of papers to be manually analyzed in about 2.5 times.

Hence, we envision that automated approaches could significantly help to reduce the SLR update effort and put forward that investigations in this direction should be encouraged and undertaken to help the community keep SLRs up to date at the pace of the rapid increase of new evidence.

In future work, we intend to improve our prototype tool to be capable of handling multiple inputs as well as further evaluate our proposition with other ML models and SLR datasets.

## 7.1 Appendix

The tool prototype and the small-scale evaluation datasets and results are openly available at ̂https://doi.org/10.5281/zenodo.7888956.

## ACKNOWLEDGMENT

We thank the authors of the ongoing SLR update considered in our evaluation for sharing their data.

## REFERENCES

[1] Charu C Aggarwal and ChengXiang Zhai. 2012. A survey of text classification algorithms. *Mining text data* (2012), 163–222.
[2] Rand Alchokr, Manoj Borkar, Sharanya Thotadarya, Gunter Saake, and Thomas Leich. 2022. Supporting Systematic Literature Reviews Using Deep-Learning-Based Language Models. In *IEEE/ACM 1st International Workshop on Natural Language-Based Software Engineering (NLBSE)*. IEEE Computer Society, 67–74.
[3] Léon Bottou. 2012. Stochastic gradient descent tricks. *Neural Networks: Tricks of the Trade: Second Edition* (2012), 421–436.
[4] Tianqi Chen, Tong He, Michael Benesty, Vadim Khotilovich, Yuan Tang, Hyunsu Cho, Kailong Chen, Rory Mitchell, Ignacio Cano, Tianyi Zhou, et al. 2015. Xgboost: extreme gradient boosting. *R package version 0.4-2* 1, 4 (2015), 1–4.
[5] CrossRef. [n.d.]. Crossref Metadata Search. https://search.crossref.org/references. Online; accessed 19 March 2023.
[6] Oscar Dieste, Anna. Griman, and Natalia Juristo. 2009. Developing search strategies for detecting relevant experiments. *Empirical Software Engineering* 14, 5 (2009), 513–539.
[7] S.C.P.F. Fabbri, K.R. Felizardo, F.C. Ferrari, E.C.M. Hernandes, F.R. Octaviano, E.Y. Nakagawa, and J.C. Maldonado. 2013. Externalising tacit knowledge of the systematic review process. *IET Software* 7, 6 (2013), 298–307.
[8] Katia Felizardo, Emilia Mendes, Marcos Kalinowski, Erica Ferreira Souza, and Nandamudi Vijaykumar. 2016. Using Forward Snowballing to update Systematic Reviews in Software Engineering. In *International Symposium on Empirical Software Engineering and Measurement (ESEM)*.
[9] Katia R. Felizardo and Jeffrey C. Carver. 2020. *Automating Systematic Literature Review*. Springer International Publishing, Cham, 327–355.
[10] Katia R. Felizardo, Erica F. de Souza, Tamiris Malacrida, Bianca M. Napoleão, Fabio Petrillo, Sylvain Hallé, Nandamudi L. Vijaykumar, and Elisa Y. Nakagawa. 2020. Knowledge Management for Promoting Update of Systematic Literature Reviews: An Experience Report. In *2020 46th Euromicro Conference on Software Engineering and Advanced Applications (SEAA)*. 471–478.
[11] Katia R. Felizardo, Elisa Y. Nakwgawa, Stephen. MacDonell, and Jose Carlos Maldonado. 2014. A Visual Analysis Approach to Update Systematic Reviews. In *International Conference on Evaluation and Assessment in Software Engineering (EASE)*. ACM, 1–10.
[12] David W Hosmer Jr, Stanley Lemeshow, and Rodney X Sturdivant. 2013. *Applied logistic regression*. Vol. 398. John Wiley & Sons.
[13] Jacob Devlin Ming-Wei Chang Kenton and Lee Kristina Toutanova. 2019. Bert: Pre-training of deep bidirectional transformers for language understanding. In *Proceedings of naacL-HLT*, Vol. 1. 2.
[14] Ashraf M Kibriya, Eibe Frank, Bernhard Pfahringer, and Geoffrey Holmes. 2005. Multinomial naive bayes for text categorization revisited. In *AI 2004: Advances in Artificial Intelligence: 17th Australian Joint Conference on Artificial Intelligence, Cairns, Australia, December 4-6, 2004. Proceedings 17*. Springer, 488–499.
[15] Barbara A. Kitchenham, David Budgen, and Pearl Brereton. 2015. *Evidence-Based Software Engineering and Systematic Reviews*. Chapman & Hall/CRC.
[16] Joseph Lilleberg, Yun Zhu, and Yanqing Zhang. 2015. Support vector machines and word2vec for text classification with semantic features. In *2015 IEEE 14th International Conference on Cognitive Informatics & Cognitive Computing (ICCI* CC)*. IEEE, 136–140.
[17] Emilia Mendes, Claes Wohlin, Katia Felizardo, and Marcos Kalinowski. 2020. When to update systematic literature reviews in software engineering. *Journal of Systems and Software* 167 (2020).
[18] Achraf Merzouki. [n.d.]. What's new in python 3.8. https://docs.python.org/3/whatsnew/3.8.html. Online; accessed 19 March 2023.
[19] Bianca M. Napoleão, Katia R. Felizardo, Erica F. de Souza, Fabio Petrillo, Sylvain Hallé, Nandamudi L. Vijaykumar, and Elisa Y. Nakagawa. 2021. Establishing a Search String to Detect Secondary Studies in Software Engineering. In *2021 47th Euromicro Conference on Software Engineering and Advanced Applications (SEAA)*. 9–16.
[20] Bianca M. Napoleão, Fabio Petrillo, and Sylvain Hallé. 2021. Automated Support for Searching and Selecting Evidence in Software Engineering: A Cross-domain Systematic Mapping. In *47th Euromicro Conference on Software Engineering and Advanced Applications (SEAA)*.
[21] NLTK Team. [n.d.]. Natural Language Toolkit. https://pypi.org/project/nltk. Online; accessed 19 March 2023.
[22] Leif E Peterson. 2009. K-nearest neighbor. *Scholarpedia* 4, 2 (2009), 1883.
[23] Joël Plisson, Nada Lavrac, Dunja Mladenic, et al. 2004. A rule based approach to word lemmatization. In *Proceedings of IS*, Vol. 3. 83–86.
[24] Python Team. [n.d.]. Urllib package documentation. https://docs.python.org/3/library/urllib.html. Online; accessed 19 March 2023.
[25] Susmita Ray. 2019. A Quick Review of Machine Learning Algorithms. In *International Conference on Machine Learning, Big Data, Cloud and Parallel Computing (COMITCon)*. IEEE Computer Society, 35–39.
[26] Nils Reimers and Iryna Gurevych. 2019. Sentence-BERT: Sentence Embeddings using Siamese BERT-Networks. In *Conference on Empirical Methods in Natural Language Processing and the 9th International Joint Conference on Natural Language Processing (EMNLP-IJCNLP)*. 3982–3992.
[27] ResearchGate. [n.d.]. https://www.researchgate.net. Online; accessed 19 March 2023.
[28] Per Runeson, Martin Host, and Austen Rainer. 2012. *Case Study Research in Software Engineering: Guidelines and Examples*.
[29] Scikit-Learn. [n.d.]. Scikit-Learn Documentation. https://scikit-learn.org/stable. Online; accessed 19 March 2023.
[30] Semantic Scholar. [n.d.]. Semantic scholar - academic graph API. https://api.semanticscholar.org/api-docs. Online; accessed 19 March 2023.
[31] Klaas-Jan Stol and Brian Fitzgerald. 2015. A Holistic Overview of Software Engineering Research Strategies. In *CESI*. IEEE Press, 47–54.
[32] Hanna M Wallach. 2006. Topic modeling: beyond bag-of-words. In *Proceedings of the 23rd international conference on Machine learning*. 977–984.
[33] Willian Massami Watanabe, Katia Romero Felizardo, Arnaldo Candido, Erica Ferreira de Souza, João Ede de Campos Neto, and Nandamudi Lankalapalli Vijaykumar. 2020. Reducing efforts of software engineering systematic literature reviews updates using text classification. *Information and Software Technology* 128 (2020).
[34] Claes. 2014. A Snowballing Procedure for Systematic Literature Studies and a Replication. In *International Conference on Evaluation and Assessment in Software Engineering (EASE)*. 321–330.
[35] Claes Wohlin, Marcos Kalinowski, Katia Romero Felizardo, and Emilia Mendes. 2022. Successful combination of database search and snowballing for identification of primary studies in systematic literature studies. *Information and Software Technology* 147 (2022), 106908.
[36] Claes Wohlin, Emilia Mendes, Katia Romero Felizardo, and Marcos Kalinowski. 2020. Guidelines for the search strategy to update systematic literature reviews in software engineering. *Information and Software Technology* 127 (2020), 106366.
[37] Claes Wohlin and Austen Rainer. 2022. Is it a case study? A critical analysis and guidance. *Journal of Systems and Software* 192 (2022), 111395.
[38] Li Zhang, Jia-Hao Tian, Jing Jiang, Yi Liu, Meng-Yuan Pu, and Tao Yue. 2018. Empirical Research in Software Engineering — A Literature Survey. *Journal of Computer Science and Technology* 33 (2018), 876–899.
[39] Barret Zoph, Colin Raffel, Dale Schuurmans, Dani Yogatama, Denny Zhou, Don Metzler, Ed H. Chi, Jason Wei, Jeff Dean, Liam B. Fedus, Maarten Paul Bosma, Oriol Vinyals, Percy Liang, Sebastian Borgeaud, Tatsunori B. Hashimoto, and Yi Tay. 2022. Emergent abilities of large language models. *Transactions on Machine Learning Research* (2022), 1–30.